\documentclass[
twocolumn,
groupedaddress,
superscriptaddress,
amsmath,
amssymb,
amsfonts, 
amstext,
]{revtex4-2}

\usepackage{dcolumn}
\usepackage{bm}
\usepackage{soul}
\usepackage[flushleft]{threeparttable}
\usepackage{url}
\usepackage{mathrsfs}
\usepackage{braket}
\usepackage{multirow}
\usepackage{array}
\usepackage{booktabs}
\usepackage{float}
\usepackage{lineno}
\usepackage{hyperref}
\usepackage{mathtools}
\usepackage{xcolor}
\usepackage{enumitem}
\usepackage{tikz}

\hypersetup{colorlinks=true, urlcolor=blue, citecolor=blue, linkcolor=blue, pdfborder={0 0 0}}

\begin{document}

\title{
Nonlinear Transport Signatures of Hidden Symmetry Breaking in a Weyl Altermagnet
}

\author{Yufei Zhao}
\affiliation{Department of Condensed Matter Physics, Weizmann Institute of Science, Rehovot 7610001, Israel}
\author{Zhiqiang Mao}
\affiliation{Department of Physics, The Pennsylvania State University, University Park, Pennsylvania 16802, USA}
\affiliation{2D Crystal Consortium, Materials Research Institute, The Pennsylvania State University, University Park, Pennsylvania 16802, USA}
\affiliation{Department of Materials Science and Engineering, The Pennsylvania State University, University Park, PA 16802, USA}
\author{Binghai Yan}
\email{binghai.yan@psu.edu}
\email{binghai.yan@weizmann.ac.il}
\affiliation{Department of Condensed Matter Physics, Weizmann Institute of Science, Rehovot 7610001, Israel}
\affiliation{Department of Physics, The Pennsylvania State University, University Park, Pennsylvania 16802, USA}

\date{\today}

\begin{abstract}
Phase transitions in solids are often accompanied by structural changes, but subtle lattice distortions can remain hidden from conventional crystallographic probes, hindering the identification of the correct order parameters. A case in point is Ca$_3$Ru$_2$O$_7$, a correlated polar ruthenate with well-characterized phase transitions, whose ground state structure has recently become a subject of debate. This uncertainty stems from extremely small atomic displacements ($\sim$ 0.001 \AA) between competing phases, beyond the resolution of X-ray diffraction, neutron scattering, or optical second-harmonic generation. In this work, we propose a method to detect hidden symmetry breaking by leveraging nonlinear transport induced by quantum geometry. We show that Ca$_3$Ru$_2$O$_7$ is a Weyl chain semimetal in both phases. The low-symmetry phase, classified as an altermagnet by symmetry, features distorted topological surface states that are asymmetric along the polar ($b$) axis. However, the nonrelativistic spin splitting is too weak ($\sim$ 0.1 meV) to be resolved directly, regarding the altermagnetism. In contrast, Weyl chains generate a large quantum metric at the Fermi surface, leading to nonlinear conductivities that are orders of magnitude stronger in the low-symmetry phase. A longitudinal nonlinear conductivity along the polar axis emerges exclusively in this phase, providing a sensitive probe to qualitatively distinguish it from the high-symmetry structure and demonstrate the emergence of altermagnetism, which is confirmed by a recent experiment. Our work establishes a route for identifying hidden symmetry breaking in complex quantum materials through the interplay of crystal symmetry, topology, and nonlinear quantum transport.
\end{abstract}

\maketitle

\section{Introduction}Hidden order refers to an ordered phase in a material (e.g., URu$_2$Si$_2$) where the underlying order parameter is not directly observable with conventional experimental methods \cite{chakravarty2001hidden, mydosh2011colloquium, okazaki2011rotational, aeppli2020hidden, soh2024spectroscopic}. While the related system exhibits a phase transition and develops some kind of internal ordering, the symmetry-breaking aspect or degrees of freedom involved remains ``hidden'' from standard probes such as x-ray diffraction or neutron scattering. 

The correlated polar ruthenate Ca$_3$Ru$_2$O$_7$ is known for its complex interplay of magnetic, electronic, and structural effects, including two antiferromagnetic (AFM) transitions (around 56 K and 48 K), and a notable metal-insulator (at 48 K) followed by an insulator-metal (at 30 K) transition \cite{Cao1997, cao2003quantum, YoshidaPRB2004, Cao2005, BohnenbuckPRB2008, BaoPRL2008, kikugawa2010}. Despite these well-characterized transitions, the precise nature of the ground state order below 30 K remains ``hidden''. For the ground state, an optical measurement \cite{LeePRL2007} indicated a possible charge/spin density wave with a pseudogap, suggesting additional subtle ordering in the ground state~\cite{Baumberger2006, Yuan2019, PDCK2020pnas}. However, crystallographic diffraction measurements showed only one crystal symmetry (space group $Bb2_{1}m$) from room temperature to 8 K \cite{yoshida2005crystal,wang2023strong}. 
Recent calculations proposed a slight symmetry reduction in the ground state, from space group $Bb2_{1}m$ to $Pn2_{1}a$, which might underpin the hidden order \cite{Rondinelli2020, Rosner2024}. The predicted atomic displacements are extremely small ($\sim 10^{-3}$ \AA), rendering them undetectable with conventional diffraction techniques.
Recent angle-resolved photoemission spectroscopy (ARPES) \cite{horio2021npj} observed massive Dirac bands near the Fermi energy. Although the reconstructed Fermi surface below 30 K was consistent with transport and thermodynamic experiments \cite{kikugawa2010, Xing2018PRB, cao2003quantum}, ARPES did not observe a symmetry reduction at the transition. 

To reveal hidden crystal symmetry-breaking, nonlinear response properties, such as nonlinear transport and second-harmonic generation (SHG) \cite{orenstein2021topology}, can serve as sensitive probes. However, two phases exhibit the same response symmetry and cannot be qualitatively discriminated by SHG. 
In transport, inversion symmetry breaking generally leads to a nonlinear anomalous Hall effect (NLAHE) driven by the Berry curvature dipole (BCD) \cite{sodemann2015quantum,zhang2018electrically,ma2019observation,kang2019nonlinear}. Further breaking of time-reversal symmetry ($\mathcal{T}$) can generate an additional NLAHE and nonlinear conductivity induced by the quantum metric dipole (QMD) \cite{Gao2014,Daniel2024,gao2023quantum,wang2023quantum}.
Heuristically, the Berry curvature captures the self-rotation of a wave packet, giving rise to a Hall effect, while the quantum metric reflects its geometric distortion~\cite{jiang2025revealing}, leading to second-order corrections in both the Hall response and longitudinal conductivity.
Compared to $Bb2_{1}m$, the $Pn2_{1}a$ structure breaks a combined symmetry $\tau \mathcal{T}$ by $\mathcal{T}$ and a half-lattice translation ($\tau$). Because $\tau \mathcal{T}$ is equivalent to $\mathcal{T}$ to constrain the nonlinear conductivity, we expect that $Pn2_{1}a$ will exhibit QMD-driven nonlinear conductivity while $Bb2_{1}m$ cannot. Furthermore, the interplay between massive Dirac bands and AFM may generate novel topological phases, which can enhance quantum geometric effects and make the nonlinear transport observable in experiments. 

In this paper, we propose to identify the ground state order of Ca$_3$Ru$_2$O$_7$ from topological surface states and quantum geometry-driven nonlinear transport. 
We point out that the $Bb2_{1}m$ phase is a conventional antiferromagnet while the $Pn2_{1}a$ phase is an altermagnet 
\cite{Wu2007,hayami2019momentum,Yuan2020,ma2021multifunctional,altermag2022PRX,Liu2022,Mazin2022}
due to symmetry reduction. However, the altermagnet-related spin splitting is too small ($< 1$ meV) to distinguish two phases from the band structure. 
In topology, we classify both $Bb2_{1}m$ and $Pn2_{1}a$ phases into Weyl chain semimetals, characterized by linked Weyl rings, rather than Dirac point or Weyl point semimetals. From $Bb2_{1}m$ to $Pn2_{1}a$, Weyl chains and related surface states deform from centrosymmetric to noncentrosymmetric at the Brillouin zone boundary, for which the $bc$ plane is optimal to demonstrate their differences by ARPES. Furthermore, Weyl chains generate strong quantum metric and consequently a nonlinear longitudinal conductivity along the polar axis ($b$) exclusively in the $Pn2_{1}a$ phase, providing a qualitative way to distinguish two phases. Caused by a weak structural distortion, it is surprising that $Pn2_{1}a$ shows three orders of magnitude stronger NLAHE than $Bb2_{1}m$. Our work paves a path to address the hidden symmetry breaking in complex materials by topological states and nonlinear quantum transport. 

\section{Methods}
The first-principles calculations were performed using the Vienna ab initio simulation package (\textsc{vasp}) with the projector-augmented wave method \cite{kresse1996efficient, kresse1999ultrasoft}. LDA \cite{ceperley1980ground}, and PBEsol \cite{perdew2008restoring} functionals were adopted. 
To treat the correlation effect of Ru-4$d$ electrons, the DFT+$U$ method was employed \cite{dudarev1998electron}. The kinetic energy cutoff of the plane-wave basis was set to 550 eV.
Including SOC and Coulomb repulsion $U$, all atomic structures were fully relaxed with fixed lattice constants until the force norm on every ion was reduced below 0.01 eV/\AA. In contrast, if SOC is not included, the relaxed structures failed to converge to either of these space groups but also exhibited unreasonable symmetry patterns. The maximally localized Wannier functions of Ru-4$d$ orbitals were built using \textsc{wannier90} package \cite{souza2001maximally}. 

The nonlinear transport conductivities are calculated by integrating the quantum geometry on the Fermi surface. In the $Bb2_{1}m$ phase, due to the presence of $\tau \mathcal{T}$, $\mathcal{G}_x$ and $\mathcal{G}_z$ symmetries, we only need to integrate the one-eighth of the Brillouin zone with a $k$-point mesh of $400 \times 400 \times 200$. In the $Pn2_{1}a$ phase, $\mathcal{G}_x$ and $\mathcal{G}_z$ symmetries divide the Brillouin zone into four parts, so we need to integrate a quarter of the Brillouin zone with a $k$-point mesh of $400 \times 800 \times 200$ ($k_x, k_z \in [0, 0.5], k_y \in [-0.5, 0.5]$).

\section{Results}
\subsection{Competing ground state structures}
\begin{figure}[t]
    \centering
    \includegraphics[width=\linewidth]{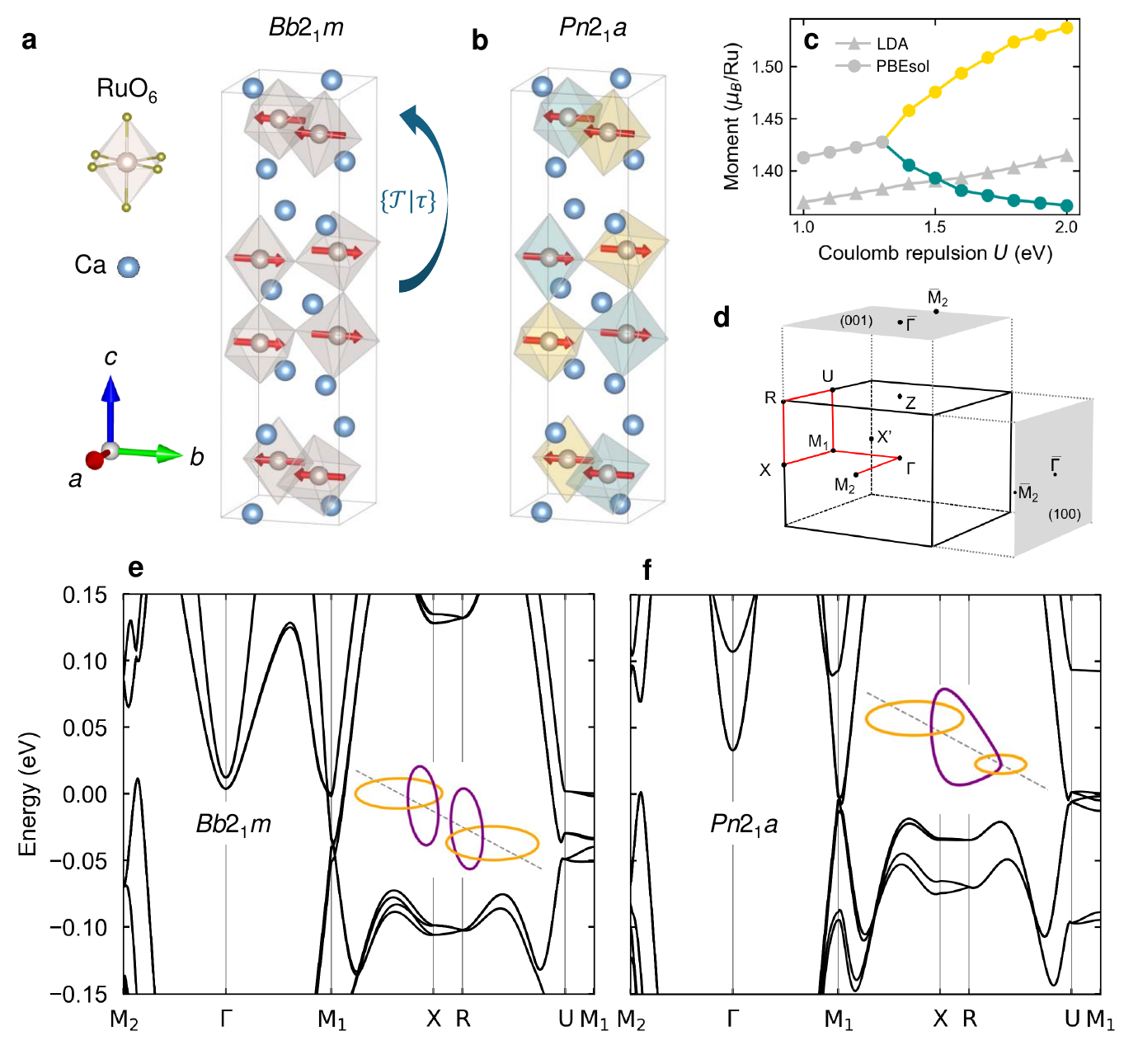}
    \caption{Crystal structure and electronic structure of Ca$_3$Ru$_2$O$_7$ with an antiferromagnetic order. (a) Unit cell in the $Bb2_{1}m$ phase with $\tau \mathcal{T}$ symmetry. (b) Unit cell in the $Pn2_{1}a$ phase without $\tau \mathcal{T}$, showing altermagnetism. Yellow and green colors denote breathing-in and -out, respectively, octahedra inside the bilayer. (c) Effect of Hubbard $U$ on the magnetic moment per Ru atom.  (d) Bulk and surface Brillouin zones. (e) Band structure of $Bb2_{1}m$ phase within LDA and $U = 2$ eV. The inset shows the symmetric Weyl chain due to crossing valence and conduction bands near the Fermi energy around the $M_1$ point.
    (f) Band structure of $Pn2_{1}a$ phase calculated with PBEsol and $U = 1.6$ eV.The inset shows the disorted Weyl chain near the $M_{1}$ point.
    }
    \label{fig1}
\end{figure}

Ca$_3$Ru$_2$O$_7$ crystallized in an orthorhombic lattice as shown in Fig. \ref{fig1}a, in which distorted RuO$_6$ octahedra form a stacked bilayer structure. 
In the $Bb2_{1}m$ phase, all RuO$_6$ octahedra are equivalent. Each bilayer exhibits a ferromagnetic order and AFM coupling to its neighboring bilayers, with magnetic moments aligning along the polar axis ($b$). Here, $a$, $b$, $c$ are lattice parameters along $x,y,z$, respectively. 
This AFM order remains the ground state in the following calculations.
Such a magnetic state includes a two-fold screw-rotation $\{C_{2y} | (0, \frac{b}{2}, 0) \}$, two glide planes $\mathcal{G}_{x}=\{ m_{x} | (\frac{a}{2}, \frac{b}{2}, \frac{c}{2}) \}$ and $\mathcal{G}_{z}=\{ m_{z} | (\frac{a}{2}, 0, \frac{c}{2}) \}$, and time-reversal symmetry combined with a half lattice translation $\tau \mathcal{T}$ = $\{ \mathcal{T} | (\frac{a}{2}, 0, \frac{c}{2}) \}$. $\tau \mathcal{T}$ connects neighboring bilayers of opposite spin sublattices. 
In the $Pn2_{1}a$ phase, however, RuO$_6$ octahedra split into two groups due to a volume-breathing distortion inside the bilayer, as shown in Fig. \ref{fig1}b. Because neighboring bilayers cannot be transformed by $\tau \mathcal{T}$ in this case, $Pn2_{1}a$ turns from an AFM into an altermagnet. Inside the bilayer, breathing distortions are literally quite weak with atomic displacements $\sim 10^{-3}$ \AA~ but lead to magnetic moment changes by $\sim 0.1$ $\mu_B$ (Fig. \ref{fig1}c).
Such weak structural distortions will modify topological bands and resultant nonlinear responses, as discussed later.

Earlier density-functional theory (DFT) calculations \cite{Rondinelli2020, Rosner2024} showed that two structural phases can be obtained by different functionals with Coulomb repulsion $U$. In structure optimizations, local density approximation (LDA) leads to the higher-symmetry $Bb2_{1}m$ phase while Perdew-Burke-Ernzerhof revised for solids (PBEsol) gives $Pn2_{1}a$ as increasing $U$ in which a structural transition happens when $U >$  1.3 eV. We successfully reproduced this trend in Fig. \ref{fig1}c.

\begin{figure*}[t]
    \centering
    \includegraphics[width=\linewidth]{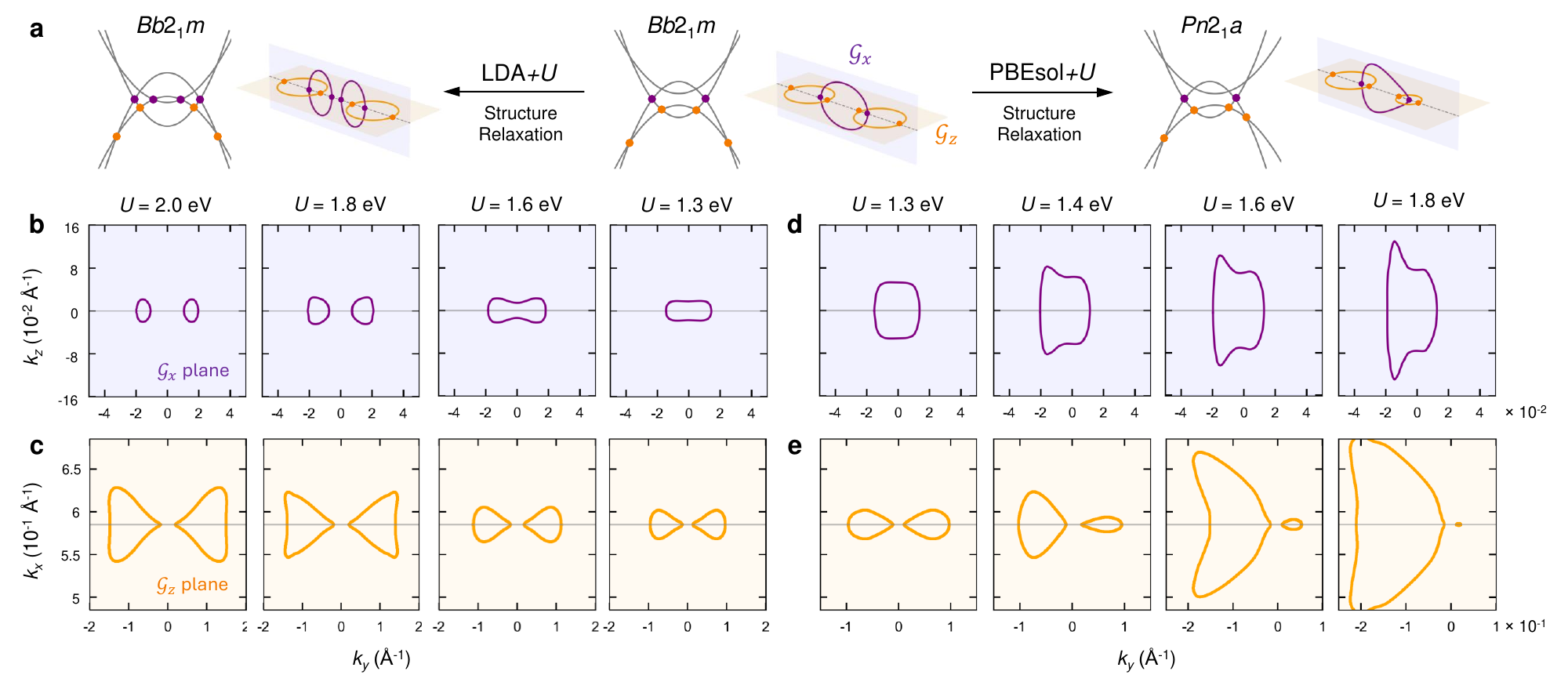}
    \caption{Evolution of the Weyl chain with respect to the Hubbard $U$. (a) Schematic of the Weyl chain structure and band structure near $M_1$. The glide planes $\mathcal{G}_{x}$ (purple) and $\mathcal{G}_{z}$ (orange) are preserved.       Band crossings between HOB and LUB are marked by purple dots (protected by $\mathcal{G}_{x}$) while the crossings between HOB and the second HOB (protected by $\mathcal{G}_{z}$) are marked by orange dots. 
    (b-e) show the evolution of Weyl rings on $\mathcal{G}_{x}$ and $\mathcal{G}_{z}$ planes. 
    (b-c) and (d-e) correspond to $Bb2_{1}m$ and $Pn2_{1}a$, respectively. }
    \label{fig2}
\end{figure*}

\subsection{Bulk band structure and Weyl chains}
Figure \ref{fig1}e, \ref{fig1}f show electronic band structures of $Bb2_{1}m$ and $Pn2_{1}a$ phases with spin-orbit coupling (SOC). In both phases, nearly linear band crossings emerge near the $M_{1}$ point, in agreement with previous theoretical calculations and ARPES measurements \cite{Rondinelli2020, Rosner2024,horio2021npj}. 
These bands exhibit spin splitting due to SOC because of the combined parity-time symmetry-breaking, lifting the degeneracy of Dirac points. One can find Weyl-like band-crossings near the $M_1$ point in Fig. \ref{fig4}. 
Inside two glide planes $\mathcal{G}_{x}$ and $\mathcal{G}_{z}$, the Weyl nodes extend to continuous rings rather than isolated nodal points. 
Here, the glide mirror symmetry protects these Weyl nodal rings but forbids the existence of isolated Weyl points, i.e., monopoles. 
For instance, we observe an hourglass-shaped band crossing along $U$-$M_{1}$ between the highest occupied band (HOB) and the lowest unoccupied band (LUB) in the $Pn2_{1}a$ phase (see Fig.~\ref{fig1}f), as part of a nodal ring on the $\mathcal{G}_{x}$ ($k_{x} = 0.5$) plane. 
On the $\mathcal{G}_{z}$ ($k_z = 0$) plane, band crossings between HOB and the second HOB form two more nodal rings. In both $Pn2_{1}a$ and $Bb2_{1}m$ phases, Weyl rings from $\mathcal{G}_{x}$ and $\mathcal{G}_{z}$ interconnect with each other and constitute a Weyl chain structure near the $M_1$ point (inset of Fig. \ref{fig1}e, \ref{fig1}f). 

Regarding the altermagnetism, we examine the non-relativistic band structure without SOC, as shown in  Supplementary Information (SI) \cite{SI} Fig. S1. According to symmetry, the $Pn2_{1}a$ phase exhibit $d$-wave-like ($d_{xz}$) spin splitting in the momentum space. However, the nonrelativistic spin splitting near the Fermi level ($E_{F}$) is less than 1 meV, much smaller than the SOC split. 
In addition, $Pn2_{1}a$ exhibits zero anomalous Hall effect due to two glide mirror symmetries. Hence, it will be hard to identify differences between $Pn2_{1}a$ and $Bb2_{1}m$ merely from altermagnetism-related spin splitting and anomalous Hall effect. 

The key distinction between the two phases manifests in their Weyl chain morphology, as illustrated in Fig. \ref{fig2}a and a model Hamiltonian below, indicating the close interplay between correlation, lattice and topology. 
In Fig. \ref{fig2}b-\ref{fig2}e, we show the Coulomb $U$ dependence of the nodal rings on two glide planes. While the existence of Weyl chains is robust against different functionals and $U$, 
their shapes are sensitive to computational parameters and symmetry breaking. When  $U$  is small ($ \le 1.3$ eV), $Bb2_{1}m$ is the preferred ground state structure with two Weyl rings on the $\mathcal{G}_{x}$ linked by another Weyl ring on the $\mathcal{G}_{z}$ plane [middle panel in Fig. \ref{fig2}a]. For $Bb2_{1}m$ phase, the increasing $U$  enhances the exchange coupling as indicated by the enhanced spin moment in Fig. \ref{fig1}c, leading to a further overlap of the conduction and valence bands. 
As schematically shown in the left panel of Fig. \ref{fig2}a, when $U > 1.7$ eV, the $\mathcal{G}_{x}$ ring splits into two rings while $\mathcal{G}_{z}$ rings expand in size (Fig. \ref{fig2}b, \ref{fig2}c). On the other hand, in the $Pn2_{1}a$ phase, the whole Weyl chain is distorted along the polar axis $b$ to be noncentrosymmetric when increasing $U$. On the $\mathcal{G}_{z}$ plane, for example, one ring expands in size while the other shrinks (Fig. \ref{fig2}e). 
We note that increasing $U$ leads to continual octahedral distortion, which changes the Weyl chain morphology sensitively and differentiates the two phases. 

\subsection{Model Hamiltonian}
We can describe the complicated evolution of Weyl chains in two phases by a simple effective model,
\begin{equation}
\begin{aligned}
    \mathcal{H}(k) = &(\textbf{k}^{2} - \mathcal{M})\tau_{z} + k_{x}\tau_{x} + 
    k_{z} \sigma_{y} (\tau_{x} + \tau_{y}) \\
    &+ (m - \eta k_{y} - k_{y}^{2})\sigma_{z}\tau_{z},
\end{aligned}
    \label{eq1}
\end{equation}
where $\tau_{x,y,z}, \sigma_{x,y,z}$ are Pauli matrices acting on orbital and spin degrees of freedom, respectively. 
$m$ is the spin splitting, $\mathcal{M}$ corresponds to the band inversion, and $\eta$ is the $\tau \mathcal{T}$ symmetry-breaking term. Controlling $\mathcal{M}$ and $\eta$ can reproduce the Weyl chain and band structure evolution in Fig. \ref{fig2}a. 

($i$) $Bb2_{1}m$ phase at small $U$ for $ -m < \mathcal{M} < m$ and $\eta = 0$: two occupied (unoccupied) bands cross each other and form two Weyl rings on $\mathcal{G}_{z}$ plane (orange) while the HOB and LUB cross each other and form a Weyl ring on $\mathcal{G}_{x}$ plane. Due to $\tau \mathcal{T}$ symmetry, the linked Weyl chain remains centrosymmetric to $M_{1}$. 

($ii$) $Bb2_{1}m$ phase at large $U$ for $ \mathcal{M} > m$ and $\eta = 0$: the nodal ring on $\mathcal{G}_{z}$ plane further splits into two loops because the second LUB crosses with the second HOB. 

($iii$) $Pn2_{1}a$ phase for $ -m < \mathcal{M} < m$ and $\eta \neq 0$: similar as case ($i$) but band structure is deformed along $k_{y}$ direction due to $\tau \mathcal{T}$ breaking.

If not specified, the $Bb2_{1}m$ and $Pn2_{1}a$ phases refer to the bulk band structures with $U$ and functionals shown in Figs. \ref{fig1}e and \ref{fig1}f, respectively, in the following surface state and nonlinear transport discussions.

\subsection{Topological surface states}
\begin{figure}[t]
    \centering
    \includegraphics[width=\linewidth]{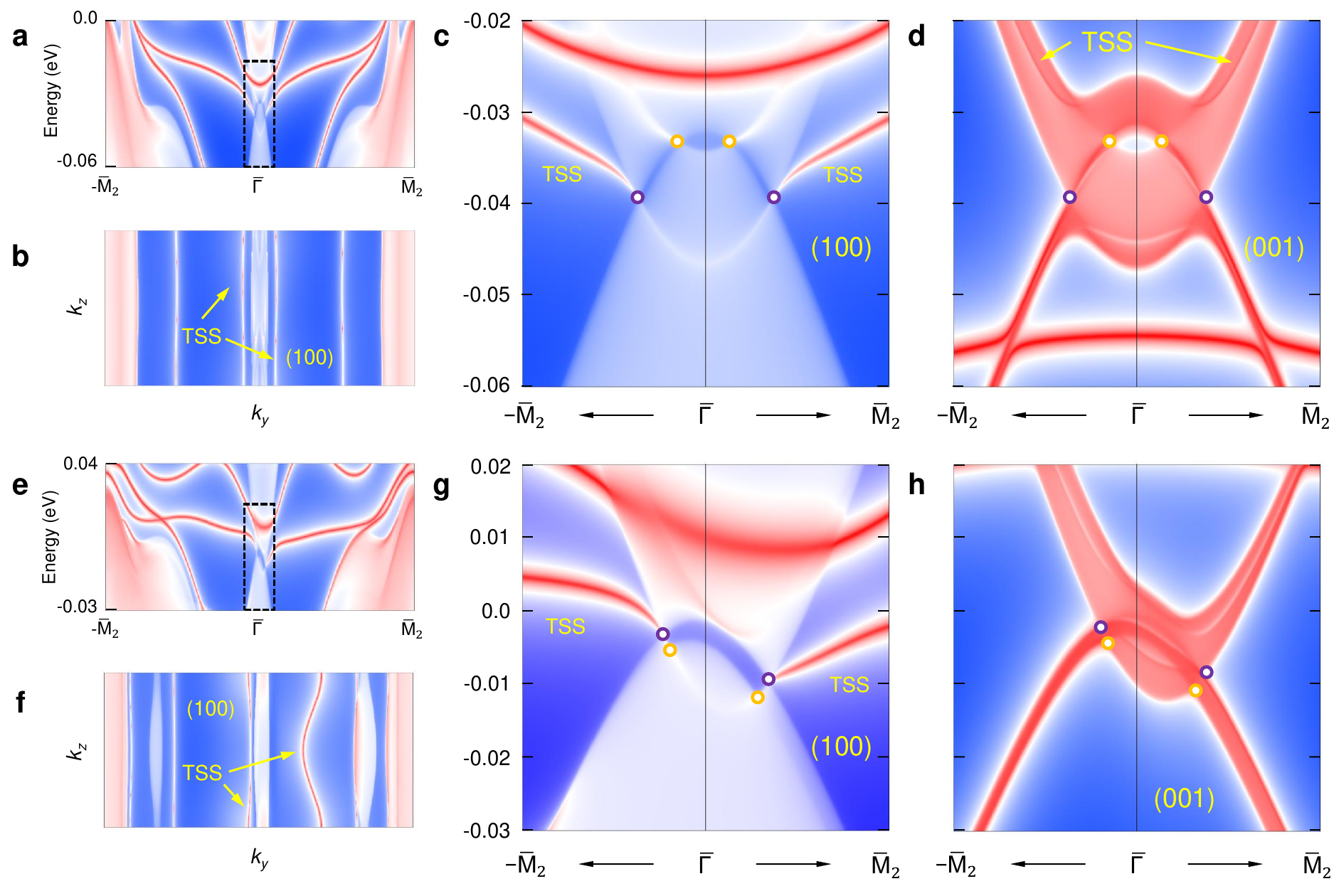}
    \caption{Topological surface states (TSSs). (a) Energy spectrum along $k_{y}$ direction on the (100) surface and (b) corresponding Fermi surface contour at $E = -0.03$ eV for the $Bb2_{1}m$ phase. (c) Zoomed-in view of TSSs. Weyl nodes from $\mathcal{G}_{x}$ ring and $\mathcal{G}_{z}$ ring, are marked as purple and orange circles, respectively. (d) Energy spectrum on the (001) surface. (e-h) Same as (a-d) but in $Pn2_{1}a$ phase. The Fermi surface contour in (f) is plotted at $E$ = $E_{F}$.}
    \label{fig3}
\end{figure}

The Weyl nodal rings on $\mathcal{G}_{x}$/$\mathcal{G}_{z}$ plane give rise to drumhead-type topological surface states (TSSs) on the (100) and (001) surfaces, which originate in the multi-band nature of Weyl nodal rings and amplify the symmetry-breaking between two phases.
On the (100) surfaces of both phases, surface band structures (see Fig. \ref{fig3}) show that Weyl nodes from $\mathcal{G}_{x}$ ring (purple circles) generate surface bands, while the $\mathcal{G}_{z}$ ring is reduced to a one-dimensional (1D) line and related TSSs are embedded inside bulk states. 
In the Fermi surface contours (Figs. \ref{fig3}b, \ref{fig3}f), these drumhead TSSs extend along the $k_{z}$ direction.
They are centrosymmetric to $\overline{\Gamma}$ point in the $Bb2_{1}m$ phase but are dramatically asymmetric in the $Pn2_{1}a$ phase, offering a strong spectroscopic signature for distinguishing them. On the (001) surface, measured by previous ARPES, however, the Weyl ring-induced TSSs remain obscured inside the bulk states for both phases (Figs. \ref{fig3}d and \ref{fig3}h). Thus, we call for ARPES measurements on the (100) surface to detect TSSs and discriminate two phases. But it might be challenging to cleave the side surface of a layered crystal.

\subsection{Nonlinear transport}
Topological features in the band structure, e.g., band crossings such as Weyl points or rings and band anti-crossing, commonly lead to giant quantum geometry (i.e., Berry curvature and quantum metric)~\cite{jiang2025revealing}. Inspired by the existence of Weyl chains, we study the nonlinear transport as a useful tool to distinguish two competing structures. 
The second-order nonlinear conductivity ($\sigma_{i;jl}$) is described by 
$j_{i}^{2\omega} = \sigma_{i;jl} E_{j}^{\omega}E_{l}^{\omega}$, 
where $E_{j}^{\omega}$ is the external ac electric field with frequency $\omega$ and $j_{i}^{2\omega}$ is the induced second-order current with $i,j,l = a,b,c$. The nonlinear conductivity is estimated by \cite{Daniel2024},
\begin{equation}
\label{eq3}
\begin{aligned}
    \sigma_{i;jl} = & -\frac{e^3 \tau_0^2}{\hbar^3}\sum_n \int_{\bm{k}} f_n \partial_{i} \partial_{j} \partial_{l} \varepsilon_{n} ~  
    \\
    & -\frac{e^3 \tau_0}{\hbar^2}\sum_n \int_{\bm{k}} f_n ( \partial_{j} \Omega^{li}_n + \partial_{l} \Omega^{ji}_n)
    \\ 
    & -\frac{e^3}{\hbar} \sum_n  \int_{\bm{k}} f_n \left( 2\partial_{i}G_n^{jl} - \tfrac{1}{2} (\partial_{j} G_n^{li} +\partial_{l}G_n^{ji}) \right) ,
\end{aligned}
\end{equation}
where
\begin{align} 
\Omega^{li}_n &= -2 \mathrm{Im}\sum_{m \neq n} \mathcal{A}_{l}^{nm}\mathcal{A}_{i}^{mn}, \label{eq4} \\
G^{li}_{n} &= 2 {\rm Re} \sum_{m \neq n}  \frac{
\mathcal{A}_{l}^{nm}\mathcal{A}_{i}^{mn}}{(\varepsilon_n -\varepsilon_m)}, \label{eq5}
\end{align}
$\partial_i \equiv \partial_{k_i}$, $\mathcal{A}_{l}^{nm} = \langle u_n  | i\partial_{l} | u_m \rangle$ is the Berry connection, $\Omega^{li}_n$ is the Berry curvature, $G^{jl}_{n}$ is the band-normalized quantum metric, and $f_{n}$ is the Fermi-Dirac function regarding the band $\varepsilon_n$. The relaxation time $\tau_0 = 58$ fs is adopted by extracting the carrier mobilities and densities from the magnetotransport experiment \cite{Mao2025}.
Because the quantum metric has an extra energy denominator in Eq.~\ref{eq5}, the quantum metric can be much more efficiently enhanced by the gapless or small-gap regions than the Berry curvature (Eq.~\ref{eq4}). 

\begin{figure}[t]
    \centering
    \includegraphics[width=\linewidth]{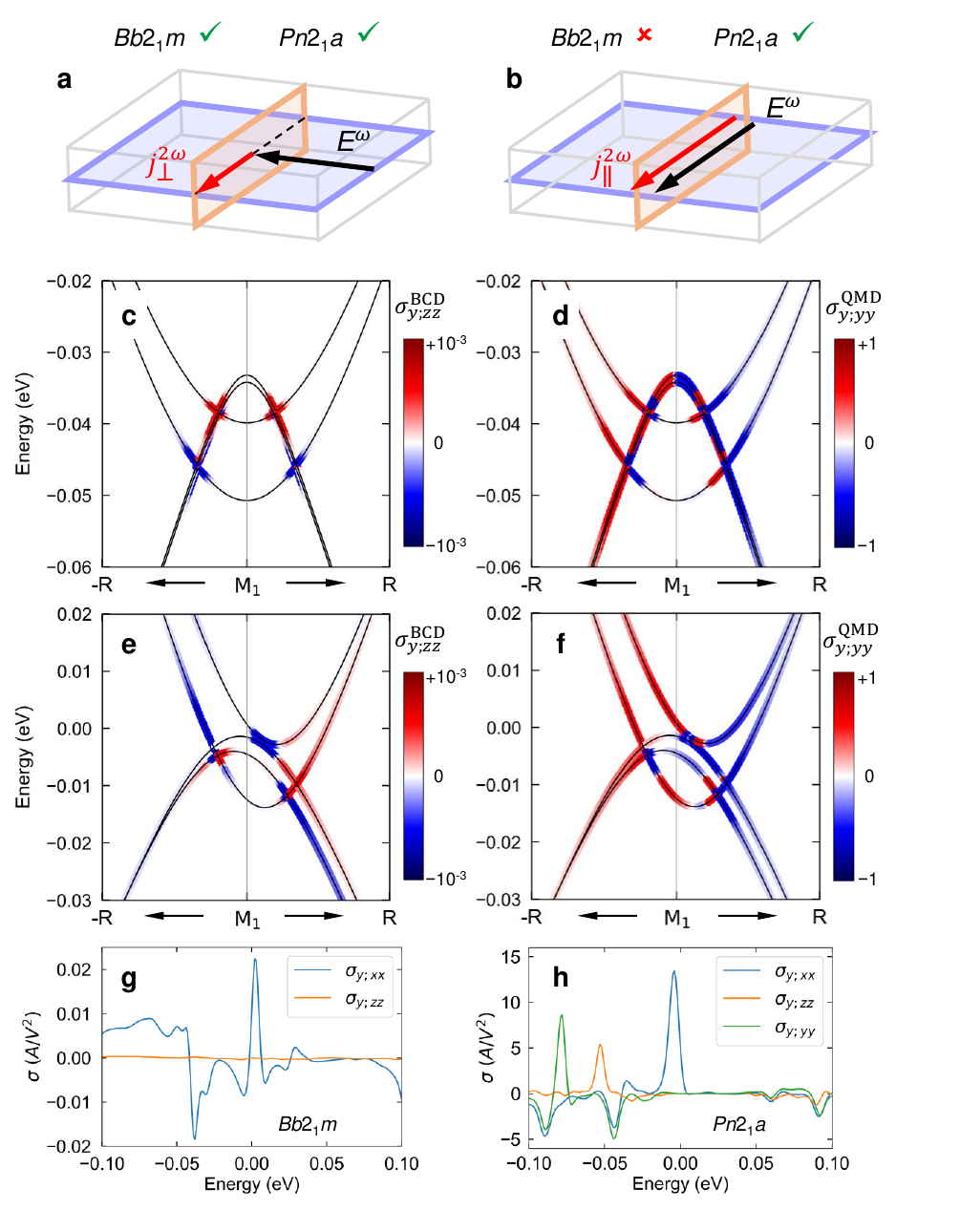}
    \caption{Nonlinear transport. (a,b) Schematics of the second harmonic generation for transverse and longitudinal transport, where the electric field $E^\omega$ generates a nonlinear current $j^{2\omega}$. The dashed line denotes the polar axis $b$. (c-f) Band structure with the projected conductivity contributed by Berry curvature dipole (BCD) and quantum metric dipole (QMD) in the $Bb2_{1}m$ phase (c,d) and $Pn2_{1}a$ phase (e,f). (g,h) The nonlinear conductivity as a function of chemical potential.}
    \label{fig4}
\end{figure}

Three terms in Eq. \ref{eq3} correspond to nonlinear Drude, BCD, and QMD, respectively, which can be classified into $\mathcal{T}$-even and $\mathcal{T}$-odd parts. The $\mathcal{T}$-odd contribution (Drude and QMD) exists only when $\mathcal{T}$ or $\tau \mathcal{T}$ symmetry is broken, generating both NLAHE and nonlinear longitudinal conductivity (see Fig. \ref{fig4}a, \ref{fig4}b). By contrast, the $\mathcal{T}$-even contribution--BCD--appears in general but only generates NLAHE. 
Therefore, the $Bb2_{1}m$ phase can only exhibit NLAHE that corresponds to $\sigma_{y;xx}$ and $\sigma_{y;zz}$ according to the magnetic space group symmetry. Here, an electric field is applied inside the glide plane ($\mathcal{G}_x$ or $\mathcal{G}_z$) and the induced nonlinear current emerges along the polar axis ($\hat{y}$).
But $Pn2_{1}a$ can present both NLAHE ($\sigma_{y;xx}$ and $\sigma_{y;zz}$) and nonlinear conductivity along the polar axis ($\sigma_{y;yy}$). 

We build the material-specific Wannier tight-binding Hamiltonian extracted from DFT calculations and then calculate nonlinear conductivity following Eq.~\ref{eq3}. As shown in Figs.~\ref{fig4}c-\ref{fig4}f, Weyl rings significantly enhance BCD and QMD in the band structure. In the $Bb2_{1}m$ phase, the $\{ \mathcal{T} | \tau \}$ symmetry enforces a symmetric distribution of BCD but an antisymmetric distribution of QMD (Fig.~\ref{fig4}c, \ref{fig4}d). Thus, the total QMD contribution ($\sigma_{y;yy}$) vanishes.  We stress that near Weyl rings, QMD-induced conductivity surpasses BCD contribution by several
orders of magnitude because the quantum metric is more efficiently enhanced by Weyl rings than Berry curvature. In the $Pn2_{1}a$ phase, a weak perturbation that breaks the QMD antisymmetry (see Fig.~\ref {fig4}f) can generate a giant net QMD, leading to around one thousand times larger nonlinear conductivities than in the $Bb2_{1}m$ phase (Fig.~\ref{fig4}g-\ref{fig4}h). 

The Weyl chain empowered nonlinear conductivities can be further demonstrated from their chemical potential dependence in Figs. \ref{fig4}g, \ref{fig4}h. In the energy range of Weyl chains, the nonlinear conductivities are generally strong with characteristic peaks. Compared to an isolated Weyl point, a Weyl ring or a Weyl chain presents more hot spots of quantum geometry and leads to stronger nonlinear conductivities
because it spans larger spaces than a Weyl point in the Brillouin zone and distributes in a broader energy window. Besides Weyl chains, generic small-gap regions in the band structure also produce large QMD. In addition, the BCD peak near the Fermi energy in the $Bb2_{1}m$ phase is caused by another band crossing (see SI Fig. S2 \cite{SI}). In the $Pn2_{1}a$ phase, the nonlinear conductivities are dominated by QMD compared to BCD and the Drude term (SI Fig. S3 \cite{SI}). 


We propose two sensitive probes to identify the subtle structural distortion and resultant band structure change.. First, the transition from $Bb2_{1}m$ to $Pn2_{1}a$ is accompanied by a significant enhancement of the nonlinear transport, a direct manifestation of the quantum metric. This can be verified by temperature-dependent transport signatures around 48 K. Second, the emergence of a second-order longitudinal conductivity along the polar axis is definitive proof for the $Pn2_{1}a$ phase, as this response is strictly forbidden in the $Bb2_{1}m$ phase.

\section{Summary}
We have predicted that Ca$_3$Ru$_2$O$_7$ is an AFM Weyl chain semimetal that hosts topological surface states and nonlinear transport. Compared to the $Bb2_{1}m$ phase, the hidden symmetry breaking in the $Pn2_{1}a$ phase drives the system into an altermagnetic state, deforms the Weyl chain structure and associated surface states, and generates a unique longitudinal nonlinear transport signal and significantly enhanced NLAHE. 
Altermagnets are usually characterized by the nonrelativistic split and anomalous Hall effect in experiments, as reported in centrosymmetric systems such as CrSb \cite{reimers2024direct,zhou2025manipulation}. 
The $Pn2_{1}a$ phase of Ca$_3$Ru$_2$O$_7$ belongs to a family of inversion-breaking altermagnet \cite{Duan2025,Gu2025}. We point out that the QMD-induced nonlinear response provides a sensitive diagnostic tool for identifying the altermagnet in polar systems, which complements other detection methods.
More broadly, our work establishes a connection between crystal geometry and quantum geometry in correlated systems. The sensitivity of nonlinear transport to hidden orders presents new opportunities to explore unknown symmetry-breaking as a new metrology beyond the limitations of conventional crystallography or spectroscopy techniques. 
It is worth noting that this method can also apply to other kinds of magnets \cite{zhao2025room} (e.g., non-collinear antiferromagnets, traditional antiferromagnets, and ferromagnets).

\textit{Note added.} A very recent experiment has measured the longitudinal and transverse nonlinear transport, as we predicted, and confirmed the $Pn2_{1}a$ phase as the ground state structure for Ca$_3$Ru$_2$O$_7$ below 30 K \cite{Mao2025}.

\begin{acknowledgments}
B.Y. and Z.Q.M. acknowledge the National Science Foundation through the Penn State Materials Research Science and Engineering Center (MRSEC) DMR 2011839. B.Y. acknowledges the financial support by the Israel Science Foundation (ISF No. 2932/21, No. 2974/23). 
\end{acknowledgments}


%

\end{document}